\documentclass[sigconf, nonacm]{acmart}

\usepackage[dvipsnames]{xcolor}
\usepackage{booktabs}
\usepackage{tabularx}
\usepackage{subcaption}
\usepackage{listings}

\AtBeginDocument{%
  }

\newcommand{\primNE}[0]{Intent Lenses} %no emph version for section headings
\newcommand{\prim}[0]{\emph{\primNE{}}}
\newcommand{\primS}[0]{\emph{Intent Lens}} %singular form

\newif\ifrevision
\revisionfalse

\ifrevision
\newcommand{\note}[1]{\textcolor{red}{#1}}

\else
\newcommand{\note}[1]{#1}

\fi

\begin{document}

%%
%% The "title" command has an optional parameter,
%% allowing the author to define a "short title" to be used in page headers.
% \title{\sys{}: Intent-Driven Synthesis and Sensemaking of Conference Experiences from Mobile Slide Captures}

\title[Intent Lenses]{Intent Lenses: Inferring Capture-Time Intent to Transform Opportunistic Photo Captures into Structured Visual Notes}

% \sys{}: Generating Intent-Mediated Visual Notes from Photo Captures of Presentations

% Intent Lenses: Generating Intent-Mediated Visual Notes from Photo Captures of Presentations

% Intent Lenses: A Sensemaking Primitive for Generating Intent-Mediated Visual Notes from Photo Captures of Presentations

% alterantives: \sys{}, NoteWeaver

%%
%% The "author" command and its associated commands are used to define
%% the authors and their affiliations.
%% Of note is the shared affiliation of the first two authors, and the
%% "authornote" and "authornotemark" commands
%% used to denote shared contribution to the research.

\author{Ashwin Ram}
\authornote{Both authors contributed equally to this research.}
\affiliation{%
  \institution{Saarland University,\\
Saarland Informatics Campus}
  \city{Saarbrücken}
  \country{Germany}
}
\email{ram@cs.uni-saarland.de}

\author{Aeneas Leon Sommer}
\authornotemark[1]
\affiliation{%
  \institution{Saarland University, \\
  Saarland Informatics Campus}
  \city{Saarbrücken}
  \country{Germany}
}
\email{aeneasleon21@gmail.com}

\author{Martin Schmitz}
\affiliation{%
  \institution{University of Koblenz}
  \city{Koblenz}
  \country{Germany}
}
\email{martin@uni-koblenz.de}

\author{Jürgen Steimle}
\affiliation{%
  \institution{Saarland University,\\
Saarland Informatics Campus}
  \city{Saarbrücken}
  \country{Germany}
}
\email{steimle@cs.uni-saarland.de}

%\author{Ben Trovato}
%\authornote{Both authors contributed equally to this research.}
%\email{trovato@corporation.com}
%\orcid{1234-5678-9012}
%%\author{G.K.M. Tobin}
%\authornotemark[1]
%\email{webmaster@marysville-ohio.com}
%\affiliation{%
%  \institution{Institute for Clarity in Documentation}
%  \city{Dublin}
%  \state{Ohio}
%  \country{USA}
%}

%%
%% By default, the full list of authors will be used in the page
%% headers. Often, this list is too long, and will overlap
%% other information printed in the page headers. This command allows
%% the author to define a more concise list
%% of authors' names for this purpose.
\renewcommand{\shortauthors}{Ram \& Sommer et al.}

%%
%% The abstract is a short summary of the work to be presented in the
%% article.
\begin{abstract}

Opportunistic photo capture (e.g., slides, exhibits, or artifacts) is a common strategy for preserving information encountered in information-rich environments for later revisitation. While fast and minimally disruptive, such photo collections rarely become meaningful notes. Existing automatic note-generation approaches provide some support but often produce generic summaries that fail to reflect what users intended to capture. We introduce \prim{}, a conceptual primitive for intent-mediated note generation and sensemaking. \prim{} reify users’ capture-time intent inferred from captured information into reusable interactive objects that encode the function to perform, the information sources to focus on, and how results are represented at an appropriate level of detail. These lenses are dynamically generated using the reasoning capabilities of large language models. To investigate this concept, we instantiate \prim{} in the context of academic conference photos and present an interactive system that infers lenses from presentation captures to generate structured visual notes on a spatial canvas. Users can further add, link, and arrange lenses across captures to support exploration and sensemaking. A study with nine academics showed that intent-mediated notes aligned with users’ expectations, providing effective overviews of their captures while facilitating deeper sensemaking. 

\end{abstract}

% \received{20 February 2007}
% \received[revised]{12 March 2009}
% \received[accepted]{5 June 2009}

\begin{teaserfigure}
    \centering
    \includegraphics[width=0.9\textwidth]{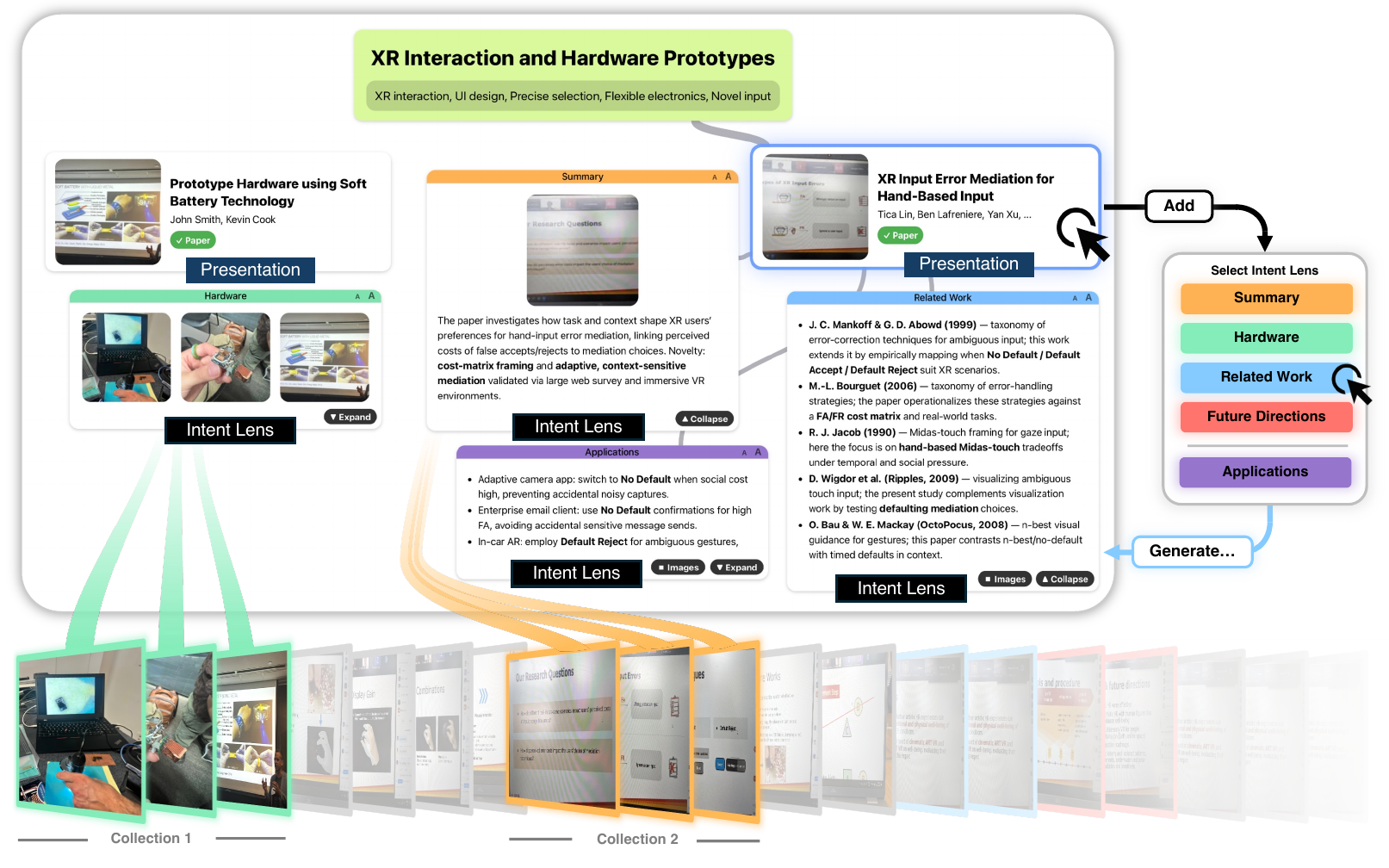}
    \caption{\prim{} are reusable interactive representations of users’ capture-time intent, inferred from opportunistic photo captures in information-rich settings such as academic conferences. These lenses automatically transform captured photos into structured visual notes that align with what users intended to capture. Users can further treat inferred lenses as reusable objects that can be reapplied to presentations (e.g.. generating a related work lens), extended through custom lenses (applications lens), and spatially arranged to support note generation, exploration, and sensemaking across captures.
    }
    \label{fig:teaser}
\end{teaserfigure}

%%
%% This command processes the author and affiliation and title
%% information and builds the first part of the formatted document.
\maketitle

\section{Introduction}

Capturing photos has become a ubiquitous strategy for taking notes of information encountered in physical, information-rich and fast-paced environments, such as conferences, lectures, exhibitions, or trade-shows. 
Attendees of such environments must simultaneously perform demanding cognitive activities, such as listen to speakers, interpret visual content, engage in conversations, and move between sessions. Manual note-taking can divert attention from these activities, forcing a trade-off between engaging in the moment and documenting information for later use. In contrast, capturing photos of slides, posters or exhibits is fast and minimally disruptive, supporting a \emph{capture now, make sense later} workflow \cite{swearngin2021scraps,rosselli2025photocaptre}.

Despite the prevalence of this behavior, our formative study with four academics found that such photo collections rarely become usable notes that users can revisit, interpret, or synthesize into meaningful knowledge.
This reflects two intrinsic challenges of opportunistic captures: First, captured content is inherently sparse. 
A photo of a slide or exhibit lacks the narration, sequential context, and surrounding discourse that give it meaning. 
Second, the intent behind each capture (what the user found significant and worth returning to)  remains implicit in the act of taking the photo.

Recent advances in AI have contributed ways to automatically generate notes from captured content, particularly for lecture videos and online meetings \cite{shin2015visualtranscripts}, and such approaches could in principle be applied to photo collections. However, the resulting notes are often generic, focusing on describing visible content rather than capturing what users found meaningful. To address this limitation, more recent work incorporates user intent into the note generation process, either through explicit input during capture (e.g., micro-notes) ~\cite{huq2025noteeline} or through retrospective prompting and querying ~\cite{li2025omniquery}. However, in practice, users rarely articulate intent during mobile capture, as the act is fast and opportunistic, and retrospectively eliciting intents across many photos is cognitively demanding and often inaccurate.

We propose treating capture-time intent not as something that users must declare, but as something that can be \emph{inferred from} contextual captures. For example, when a person photographs a slide during a talk, they are selectively capturing something they anticipate revisiting, such as a key method, an intriguing result, or an idea for future work. Inferring these implicit intents can help generate notes that better align with what users intended to capture. \note{Furthermore, because many such intents recur across multiple captures, we argue that intent should persist beyond one-time note generation as a reusable level of interaction through which users can explore, compare, and synthesize captured information over time. In this work, we explore how such intent (and other relevant contextual data) can be inferred from sparse captures and operationalized to support more meaningful note generation and sensemaking.}

Building on this perspective, we introduce \prim{}, a conceptual primitive for reifying implicit intents inferred from contextual captures. Each intent lens encodes a particular way of engaging with the captured information by specifying the function to perform, the information sources to focus on, and how the output is visually represented at an appropriate level of detail. Applying different lenses to the same underlying data surfaces different patterns and relationships, much like optical lenses that bring certain features into focus while leaving others in the background. In this way, \prim{} act as persistent, reusable operators over data that generate content aligned with users' intent. \prim{} build on prior work that represents LLM interactions as user-specified objects ~\cite{kim2023cgl, masson2024direct}, but crucially differ in that they are dynamically generated from contextual captures, allowing interpretations to emerge from data rather than explicit user input.

To investigate and evaluate this concept concretely, we focus on a domain where opportunistic photo capture is well-established and its limitations are acutely felt: academic conference attendance. 
Based on findings from the formative user study, we developed an interactive system that infers \prim{} from conference photo captures and uses them to automatically generate intent-mediated visual notes (\autoref{fig:teaser}). The system further augments captured photos with additional context such as the corresponding digital article when available, providing richer signals for lens-based interpretation. Users can further create, link, and arrange lenses across multiple presentations, constructing evolving visual notes that support exploration, comparison, and synthesis. By reusing lenses across different presentations, users can generate consistent views of information, surface relationships across talks, and externalize their sensemaking process. We evaluated the prototype through a user study with nine academics from different scientific disciplines. Our findings show that \prim{} facilitated the generation of notes that aligned with what users expected to capture, providing meaningful overviews of their conference experiences while supporting deeper post-hoc sensemaking.

In summary, we contribute:
\begin{itemize}

\item A formative study of photo-based note-taking at academic conferences, identifying key challenges in revisiting opportunistic captures and common user intents behind them.

\item The concept of \emph{\prim{}} as a primitive for reifying implicit intent from contextual photo captures. We present a technical realization that infers capture-time intent from sparse captures using multimodal structured reasoning and operationalize it as reusable prompt-based representations.

\item An interactive system that instantiates \prim{} on a spatial canvas, enabling users to generate, apply, and compose lenses over captured artifacts to create structured, revisitable visual notes.

\item Evaluation results providing insights into how \prim{} support note generation, exploration, and sensemaking, along with design implications for future intent-mediated systems.

\end{itemize}

\section{Related Work}

\subsection{Approaches and Tools for Note-Taking}

HCI research has long sought to develop systems that support note-taking \cite{morehead2019notetaking}, given the substantial cognitive demands it places on users, who must simultaneously interpret information, select what is relevant, and reorganize it into meaningful representations \cite{piolat2005cognitive}. Early efforts aimed to assist note-taking by scaffolding capture \cite{hinckley2012gather, cao2022videosticker, roy2021highlight, kuznetsov2022fuse, qiu2025marginalia}, linking \cite{chiu1999notelook, olsen2004screencrayons, hinckley2007inkseine, subramonyam2020texsketch, qiu2025marginalia}, and organization \cite{brandl2010nicebook, tashman2011liquid, meng2016hynote, swearngin2021scraps, tsai2025gazenoter}, while leaving composition of notes largely to the user. For example, NoteLook \cite{chiu1999notelook} integrates video with ink annotations to support revisitation, and LiquidText \cite{tashman2011liquid} enables flexible excerpt manipulation and spatial reorganization to facilitate synthesis. More recent systems have considered how notes can be automatically generated \cite{shin2015visualtranscripts, li2021summary, dang2022beyondtext, xu2023autonote, zhao2025noteit, jiang2025nexanota} or expanded upon using AI. For instance, NoTeeLine \cite{huq2025noteeline} expands short user-authored micronotes into fuller notes while preserving writing style using LLMs. Across these approaches, note-taking continues to rely on textual inputs or transcripts as the primary substrate.

In contrast, in settings such as conferences and exhibitions, users increasingly rely on photo capture of slides, posters, or exhibits as a primary note-taking strategy \cite{thakur2011mobilecapture, swearngin2021scraps, rosselli2025photocaptre}. Photo capture is fast and minimally disruptive, supporting a \emph{capture-now, make sense later} workflow, but they often fail to serve the purposes of note-taking, instead accumulating into large, unstructured collections that are difficult to revisit, synthesize, or use as effective notes \cite{wong2023take}. Our work addresses these challenges by inferring the implicit intent behind photo captures and operationalizing it to generate visual notes that support overview, exploration, and reflective sensemaking over captured materials.

\subsection{Intent-Driven Systems}

With recent advances in Generative AI, there has been active interest in developing systems that provide adaptive and context-aware support for various tasks by eliciting or inferring users' intent \cite{zulfikar2024memoro, lee2024gazepointar, lee2025sensible, li2025satori, shaikh2025gum, fang2025mirai, kim2026speechless}. A first class of systems relies on users explicitly articulating their goals at interaction time \cite{dogan2024xr, li2025omniquery, gmeiner2025intenttagging}. Intent is expressed through natural language queries or multimodal prompts, after which the system resolves and executes that intent by grounding it in contextual information. For example, XR-Objects enables object-aware question answering in AR through user-driven queries \cite{dogan2024xr}, where responses are contextualized to the surrounding environment, and OmniQuery supports personal question answering over multimodal memories but requires users to formulate explicit retrieval queries \cite{li2025omniquery}. 

A second line of work moves from reactive execution to proactive inference \cite{li2024omniactions, lee2025sensible, li2025satori, pu2025promemassist, kim2026speechless}. These systems anticipate user needs by modeling context, behavior, or latent goals. For example, OmniActions \cite{li2024omniactions} predicts follow-up digital actions based on multimodal sensory input, such as images or audio. Similarly, Sensible Agent \cite{lee2025sensible} reasons over situational context to determine both what assistance to provide and how to deliver it. Satori further extends this direction by maintaining a Belief–Desire–Intention model to infer latent goals and guide proactive AR interventions \cite{li2025satori}. Such systems operate in real-time environments, where intent is inferred from continuous multimodal streams and immediately acted upon.

We build on this broader shift toward intent modeling but address a fundamentally different setting, where intent must be reconstructed retrospectively from sparse, photo captures rather than inferred from continuous multimodal data or explicit queries. We also differ in how the inferred intent is used. \note{Instead of using intent as a transient conditioning for note generation, we reify it into interactive objects that persist beyond the initial generation step as a way to interact and make sense of the captured data. This builds on the concept of \emph{reification}~\cite{beaudouin2000reification} and prior work that has showed how reifying task elements into interactive objects can facilitate workflows~\cite{beaudouin2000intstrument, xia2016objectdrawing, xia2018dataink, kim2023cgl, masson2024direct}. \prim{} extend this line of work by dynamically generating these object representations from captured data, rather than relying on user-specified task definitions.}

\subsection{Tools for Sensemaking}

An important precursor to our work is ButterflyNet \cite{yeh2006butterflynet}, which enables users to capture photos, audio, and annotations during field observations and revisit them later for organization and reflection. It frames capture as part of an iterative sensemaking loop \cite{russell1993sensemaking}, where users collect fragments opportunistically and later externalize, structure, and reinterpret them. This separation between lightweight capture and later synthesis has established what we refer to as a \emph{capture now, make sense later} workflow, which directly informs \prim{}. 

More recent AI-driven systems have advanced sensemaking by making semantic relationships explicit and interactively navigable \cite{jiang2023graphologue, suh2023sensecape, suh2025storyensemble}. For example, Sensecape externalizes levels of abstraction through hierarchical and node-link views \cite{suh2023sensecape}, allowing users to move between detailed content and higher-level conceptual groupings. Similarly, Graphologue constructs editable knowledge graphs to support exploration and sensemaking across documents \cite{jiang2023graphologue}. These systems emphasize making structure visible and navigable, enabling users to explore information non-linearly through connections, hierarchies, and abstraction layers. \prim{} build on this line of work but introduce a complementary perspective. Rather than focusing solely on externalizing structure or enabling nonlinear workflows, we bring capture-time intent into the workspace as an interactive object—something users can apply and adapt over time to surface patterns and relationships from captured data, while flexibly organizing it to support sensemaking.

\section{Formative Study}

\note{Despite the growing prevalence of opportunistic photo capture as a lightweight note-taking strategy in information-rich environments, there is limited research on how people capture, and revisit such materials in practice. Academic conferences provide a compelling setting to study this \emph{capture now, make sense later} workflow: attendees encounter dense, fast-paced information across talks, posters, demos, and exhibits, and often rely on rapid photo capture for later reflection.} Unlike more structured settings such as lectures, conferences also involve highly individualized goals, from identifying related work to gathering inspiration or building a high-level understanding of unfamiliar areas. To better understand current practices and challenges of this workflow, we conducted a formative study with conference attendees, examining how they capture, organize, and revisit conference photos.

\subsection{Participants and Procedure}

We conducted semi-structured interviews with four volunteers (2 male, 2 female, aged 24-28 years), all PhD students from the university who had attended at least one academic conference or workshop in the past five years. We focused on early-career researchers, as they frequently attend diverse talks and often rely on external note-taking to build structured understanding of research areas and communities. The interviews covered participants’ current note-taking practices, their use of photos as notes, post-conference organization and reuse, knowledge-building goals, challenges and pain points, and desired support. Participants were also asked to review images they had captured at past conferences and reflect on why they took specific photos and how (or whether) they revisited them. Interviews lasted approx. 45 minutes and were conducted either in person or remotely. With informed consent, sessions were audio recorded and transcribed for thematic analysis.

\subsection{Findings and Discussion}

\newcommand{\jquote}[1]{\emph{“#1”}}
\newcommand{\pquote}[2]{\jquote{#1} (#2)}

Across interviews, participants consistently described a shift away from manual note-taking toward photo capture, emphasizing the speed and convenience of the approach: \pquote{It’s always quick. You just take a picture of each slide and then you’re not busy typing the whole time}{P1}. Manual note-taking was often described as distracting: \pquote{I usually don’t like writing things down or typing because during a talk that can get quite distracting}{P2}. Several participants noted observing similar behavior in other conference attendees, with some attendees photographing \pquote{almost the entire presentation}{P1}. However, there were significant limitations in how those images supported later use as notes. In the following, we present the key challenges that emerged from the interviews:\\

\noindent \textbf{C1. Revisit and Retrieval are Difficult.} Although capturing slides was quick and convenient, revisiting them later was often cumbersome and inefficient. Participants described difficulties locating specific photos within large, unstructured collections, noting that \pquote{it can take time to find the one I want}{P3}, especially because \pquote{in the photo folder, the thumbnails are small and many look similar}{P3}. Photos were typically stored without organization and \pquote{just get mixed into my camera roll}{P4}, making it \pquote{very easy to lose context}{P2}. This lack of structure made retrieval challenging, with one participant admitting, \pquote{I’m the kind of person who has ... [many] pictures, but I’m afraid I can’t find it in this mess.}{P1}. As a result, participants reported rarely revisiting their conference photos, even though they had captured it with a desire to do so.\\

\noindent \textbf{C2. Lack of Overview and Structured Representation.} While photos preserved individual slides, they offered little support for forming a coherent understanding of the overall conference experience. Participants found it difficult to move beyond fragmented, slide-level captures to reconstruct broader themes, connections across sessions, or high-level takeaways. As a result, they expressed a need for representations that provide a concise overview of their experiences, with P4 noting they would like to \jquote{have an overview of what I saw and why it was relevant}. Participants also emphasized that visually structured representations could improve readability and encourage revisitation. \\

\noindent \textbf{C3. Lack of Support for Intent-Mediated Sensemaking.} Our  analysis further revealed that participants’ photo captures were often guided by recurring underlying intentions about what they found noteworthy. Across interviews, we identified common capture-time intents, such as getting a summary or noting down related work, as well as participants’ expectations for how notes supporting these intents should be represented. These recurring themes are summarized in \autoref{tab:intent_formative}. Furthermore, the number of photos taken during a presentation also reflected their level of interest: \pquote{If something is very related ... the number of photos goes up accordingly}{P4}. However, these intentions were not preserved after capture, making it difficult to recover why a photo was taken or to identify patterns across talks. As a result, retrospective sensemaking was challenging, often requiring participants to \pquote{search for that information online}{P3} to retrieve details that were not readily accessible from the captures themselves.

\begin{table}[t]
\centering
\small
\caption{Common capture-time intents and what users expected to see about them in their notes.}
\begin{tabular}{ll}
\hline
\textbf{Implicit Intent} & \textbf{Preferred Note} \\ \hline
Summary & short textual overview \\ \hline
Related work & \begin{tabular}[c]{@{}l@{}}List of paper references with \\ names, authors, published venue\end{tabular} \\ \hline
Future Directions & \begin{tabular}[c]{@{}l@{}}List of short bullet points \\ with the relevant captured photo\end{tabular} \\ \hline
Design Inspirations & Grid of pictures \\ \hline
\end{tabular}
\Description{Add later}
\label{tab:intent_formative}
\end{table}

\begin{figure*}[t]
    \centering
    \includegraphics[width=0.9\textwidth]{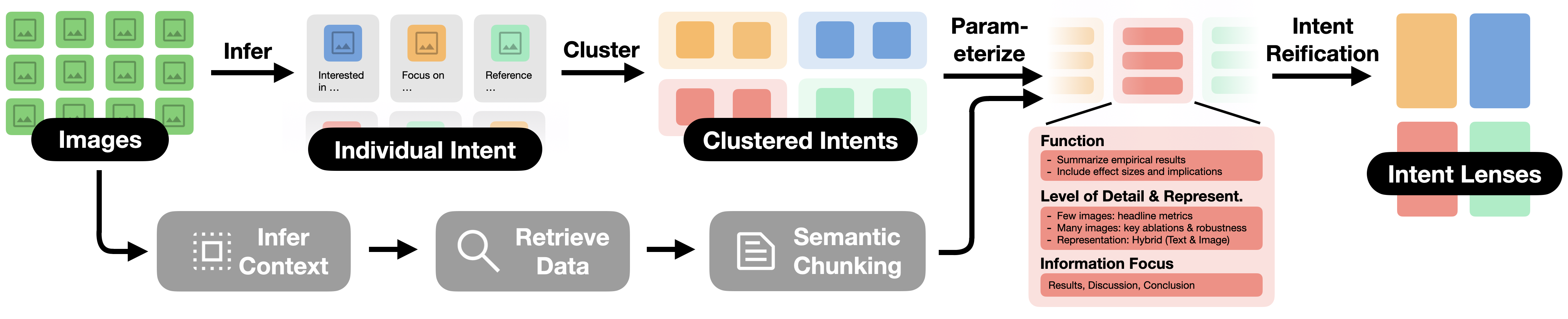}
    \caption{Intent Lens construction pipeline. It consists of four LLM-based stages: intent inference, intent clustering, intent parameterization, and intent reification. Starting from opportunistic photo captures, the pipeline infers fine-grained image-level intents, clusters them into reusable shared categories, parameterizes each intent by function, information focus, and representation, and reifies the result into reusable Intent Lenses for generating structured visual notes.}
    \label{fig:pipeline}
\end{figure*}

\subsection{Design Goals}

Based on the challenges C1–C3, we derive four design goals to guide note generation and sensemaking from photo captures. \\

\noindent \textbf{D1. Use Captured Images as the Primary Input.} 
Photo capture is fast, opportunistic, and minimally disruptive, but existing workflows make revisiting and retrieval difficult (C1). While collecting additional input from users at capture time could improve downstream note generation, it would risk making capture cumbersome. Therefore, a system should operate directly on captured images, leveraging contextual signals that can be inferred from the captures themselves. \\

\noindent \textbf{D2. Infer Implicit Capture-Time Intent.}
\note{Users capture content with diverse, implicit intentions, such as methods, inspiration, or related work, which are not preserved after capture (C3). A system should infer these latent capture-time intents and generate notes that align with this intent.} \\

\noindent \textbf{D3. Facilitate Intent-Mediated Sensemaking.}
Raw photo collections lack structure and make it difficult to form a coherent understanding of the broader experience (C2, C3). A system should therefore support sensemaking through structured notes grounded in inferred capture-time intent, enabling users to revisit captures, compare information across notes, and synthesize higher-level patterns and relationships that are not apparent when captures are viewed in isolation. \\

\noindent \textbf{D4. Make Intent Reusable Across Captures.}
\note{Our findings further suggest that many capture-time intents recur across multiple captures and continue to shape how users revisit and compare content over time (C2, C3). Rather than using inferred intent only once for note generation, a system should externalize capture-time intent as a persistent and reusable interaction object for sensemaking.}

\section{\primNE{}}
\label{sec:concept}

\note{We introduce \prim{}, a conceptual primitive that reifies implicit capture-time intent into explicit objects that transform opportunistic photo captures into structured, revisitable visual notes (D2, D4). While conditioning note generation on inferred intent can better align notes with what users intended to capture, many such intents recur across multiple captures and remain valuable beyond a single output. For example, in academic conference settings, users often wish to generate summaries across talks or compare methods and findings. We argue that reifying inferred intent into persistent interactive objects enables users to engage with captured data directly at the level of intent, supporting both intent-mediated note generation and sensemaking across captures~(D3).}

\note{Building such a reusable representation of intent from sparse opportunistic photo captures is challenging. First, the intent behind each capture remains implicit in the act of taking the photo and must therefore be inferred. Second, opportunistic captures are inherently sparse and often provide only partial context, such as isolated slides or exhibits without surrounding narration or discourse, making both intent inference and the downstream application of that intent for note generation difficult. To address this, an \primS{} defines a reusable way of engaging with captured information through three parameters: (1) the function to perform, (2) the information to focus on, and (3) the representation of the results. The function specifies the operation over the data, such as summarization or extraction; the information focus selectively attends to the most relevant signals from captured images and associated contextual sources; and the representation determines how outputs are presented, such as structured text, visual elements, or hybrids, and at what level of detail.}

An \primS{} is constructed in three key steps (\autoref{fig:pipeline}) discussed below: (1) inferring capture-time intent from individual photos, (2) clustering these into reusable intent categories, and (3) parameterizing each intent by clearly defining how it operates over the data.

\subsection{Intent Inference}

Constructing an \primS{} begins with a collection of photos captured by the user (D1). For each image, we infer the user’s implicit capture-time intent, i.e., what they found meaningful and intended to revisit (D2). To do so, we first use a vision–language model (VLM) to extract structured information from each image, including all visible text, the title, descriptions of visual elements, and an explanation of the depicted content, which is stored as JSON. This extracted information is then passed to an LLM that performs structured chain-of-thought reasoning to infer the capture-time intent (\autoref{fig:pipeline}, Infer step). The reasoning considers multiple aspects of the captured content, including the type of material depicted, its information density, the likely motivation for capturing it, and how the user might engage with it later. Based on this reasoning, the model produces a concise description that summarizes the inferred intent for each image. We use GPT-Image-1 as the initial VLM for image understanding. GPT-5 serves as the LLM for this and subsequent steps. \autoref{sec:model_infer_intent_prompt} provides the prompt for this step.

\subsubsection{Enriching Sparse Collections} Captured images often provide only partial information, which may not be sufficient to fully address the inferred intent. To address this, the VLM in this step also identifies external sources that can enrich the captures, such as digital articles or product specifications, inferred from cues available in the images including titles, references, or other identifiable elements  (\autoref{fig:pipeline}, pipeline in gray). Retrieved sources are segmented into smaller semantic chunks (e.g., typical sections in digital articles), which \prim{} can later selectively access during note generation. In addition, a summarized version of each retrieved source is included as an additional chunk. Incorporating this context enables outputs that are more complete and better aligned with the inferred intent, supporting revisitation and sensemaking (D3).

\subsection{Intent Clustering}

The inferred intent for each captured image often varies slightly, resulting in many fine-grained, per-image intent descriptions. While these capture nuanced differences in what the user found meaningful, they are not directly reusable. A core goal of \prim{} is to support reusable and consistent interaction with captured material (D4). To achieve this, we consolidate individual intents into a smaller set of shared, reusable intents via clustering. An LLM clusters per-image intent descriptions across all images into broader intents using chain-of-thought reasoning, grouping them based on the their semantic relatedness, abstraction of user goals, and coverage of all individual intents (\autoref{fig:pipeline}, Cluster step). For each cluster, the LLM generates a label and description capturing the shared intent. Subsequently, each image is assigned to the cluster that best matches its inferred intent. \autoref{sec:model_cluster_intent_prompt} shows the prompt for this step.

\subsection{Intent Parameterization}

While clustering provides a high-level description of user intent, these descriptions remain inherently underspecified. Leaving these decisions implicit forces the model to infer missing details, which can lead to inconsistent or suboptimal outputs \cite{yang2025promptsdontsayunderstanding}. To address this, we explicitly parameterize each clustered intent along three key dimensions: the intent function, the information focus, and the intent representation (\autoref{fig:pipeline}, Parameterize step). Each parameter is inferred by the LLM instance based on the label and description of the clustered intent, as well as in-context examples of how these parameters should be instantiated for a given domain. \autoref{sec:model_intent_lens_prompt} provides the prompt for this step.

\subsubsection{Intent Function}

The function parameter translates the description of the clustered intent into clear tasks that when performed on the data, transform it into a visual note that realizes the intent. For example, for a related work intent, the function specifies the task as extracting and listing the most relevant prior work.

\subsubsection{Information Focus} This parameter specifies which parts of the data, including captured photos and retrieved external sources, should be attended to when executing the intent function. Prior work shows that increasing input context does not necessarily improve performance. Instead, longer inputs often introduce irrelevant information, making it harder for models to identify and use relevant signals \cite{liu2024llmcontext, du2025contextlengthhurtsllm}. Accordingly, rather than using all photos and the full set of retrieved sources, the lens restricts the input to photos associated with the given intent and relevant chunks from external source. This targeted selection reduces noise and improves the precision of the generated outputs.

\subsubsection{Intent Representation}

The representation parameter determines how the generated visual note is presented, including format and level of detail. The LLM infers an appropriate format based on the intent, selecting between text, image, or a hybrid. In our current formulation, this choice is drawn from a fixed set; however, the parameterization is extensible, allowing the VLM to infer from arbitrary sets or dynamically generated UI descriptions \cite{cao2025malleable, min2025meridian}. For example, for a "related work" intent, the function may specify extracting relevant prior work, while the representation determines that these are presented as a bulleted list rather than continuous prose, and how many references to include. The level of detail is adapted based on the number of captured images, which our formative study identified as a proxy for user interest. To support this, the representation parameter includes heuristic guidelines that map image counts to output granularity, ensuring that intents associated with more images yield richer outputs.

\subsection{Intent Reification}

The three inferred parameters (i.e., intent function, information focus, and intent representation) are unified into a single prompt that constitutes an \primS{} (D4). Applying an \primS{} transforms captured data into notes aligned with the function encoded by that lens. They can be flexibly applied over the captured data at different levels of granularity. When applied to an individual photo capture, the \primS{} operates solely on that image, using its visual content as context to produce a note. When applied to a group of photos sharing a common context, such as images of a presentation or poster, the \primS{} generates a note by selectively attending to the relevant photos and retrieved external source chunks within its focus. In both cases, the format of the resulting note is determined by the representation parameter encoded by that lens.

\note{Taken together, \prim{} provide a general mechanism for transforming sparse, opportunistic captures into structured, intent-aligned representations that support revisitation and sensemaking. We next demonstrate how this mechanism operates in a concrete domain by showing the interactions enabled through \prim{}.}

\section{\primNE{} for Presentation Captures}
\label{sec:application}
\begin{figure*}[t]
    \centering
    \includegraphics[width=\textwidth]{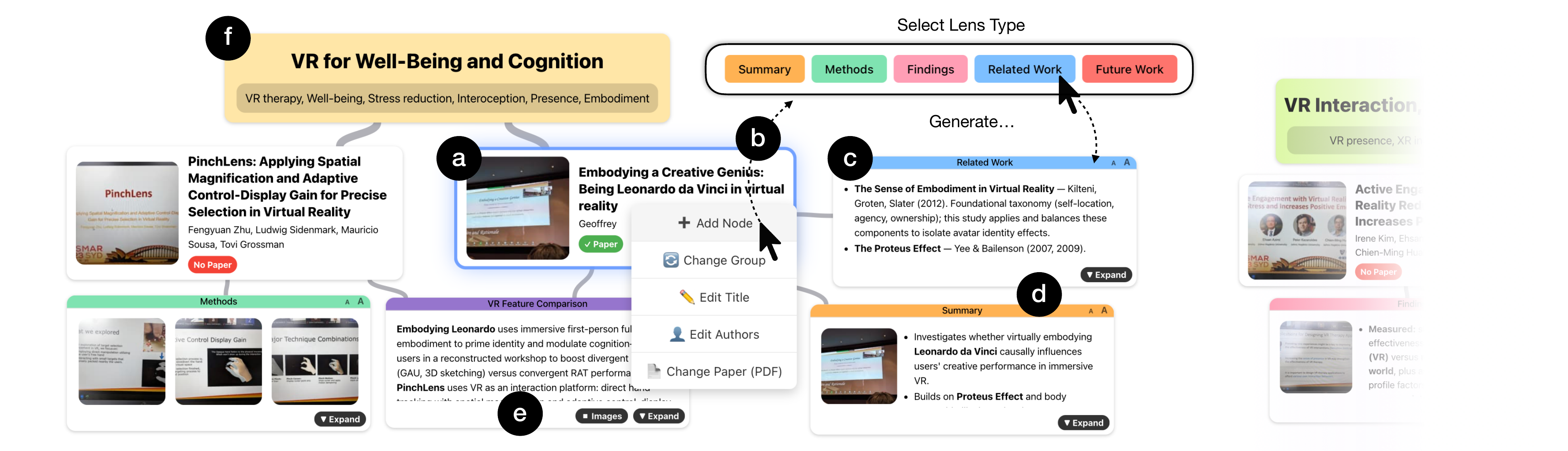}
    \caption{Graphical interface for exploring and making sense of opportunistic conference photo captures using \prim{}. Captures from a given talk or poster are represented as (a) Presentation nodes. For each Presentation, the system automatically instantiates Lens nodes for the subset of inferred \prim{} reflected in its captured images. Users can (b) reuse an inferred lens type and (c) instantiate a corresponding Lens node linked to the Presentation. (d) Multiple Lens nodes can be attached to a single Presentation node, enabling multifaceted notes that support parallel exploration and sensemaking. Users may also create (e) custom Lens nodes that synthesize information across Presentations nodes. In addition, the system uses (f) a special Lens node to surface shared themes and keywords across related papers. }
    \label{fig:interface}
\end{figure*}

To investigate how \prim{} can facilitate intent-mediated note generation and sensemaking from opportunistic photo captures, we instantiate them in the domain of academic conference photos. We develop an interactive system that infers \prim{} from the captured conference photos and automatically uses them to generate structured, revisitable visual notes that align with what users intended to capture (\autoref{fig:interface}). These lenses can then be further reused and adapted for further exploration and sensemaking. 
We describe how \prim{} are applied in this domain and the key interactions they enable for note generation, exploration, and sensemaking.

\subsection{System Overview}

The system is a zoomable infinite canvas with node-based elements (Figure~\ref{fig:interface}) implemented using ReactFlow. It supports two node types that users can freely arrange and connect: Presentations and Lenses.

Presentation nodes (Figure~\ref{fig:interface}a) represent groups of photos captured from a single talk, poster, or demo. These nodes are created from user-provided images and enriched with metadata inferred from the captures (e.g., title and associated paper). They are visualized as cards showing the presentation title, and authors, and a captured image from the group. Lens nodes represent instances of \prim{}. A Lens node can be created for a Presentation node, generating a visual, card-based note that is linked to that Presentation node (Figure~\ref{fig:interface}b) . Each Lens node (Figure~\ref{fig:interface}c) is visually encoded by both color and label, indicating the \primS{} type it belongs to. Clicking on a Lens node reveals the connected Presentation nodes.

The workflow begins with users uploading conference photos grouped into folders by presentation (D1). The pipeline (\autoref{sec:concept}) processes these images to infer capture-time intents, clusters them, and generates a set of five \prim{} specific to the user (D4). Common inferred \prim{} include summary, related work, future directions, and methods, while more specific intents may also emerge, such as design inspirations, hardware aspects, or applications. To enrich the captured photos, the system automatically retrieves associated digital articles using presentation titles inferred from the images by the VLM through a combination of the Semantic Scholar API and SerpAPI.

We next describe how these inferred lenses support note generation, exploration, and sensemaking across presentations.

\subsection{Note Generation and Sensemaking with Lenses} 

\subsubsection{Intent-Mediated Note generation}

The system automatically instantiates Lens nodes for each Presentation node, but only for the subset of inferred \prim{} reflected in its captured images. Each Lens node selectively attends to relevant photos and extracted paper sections, and applies a corresponding function, such as summarization or listing related work, to generate a structured note. For example, if photos indicating summarization are present, a "summary" lens is created and generates a note of key ideas; if no photos reflect engagement with future work, no corresponding lens or note is produced. The visual format of each Lens node is determined by the representation parameter of the corresponding \primS{}, resulting in a text-only, image-only, or hybrid card view. The level of textual detail further depends on the number and coverage of captured images, with more extensive captures producing richer and more comprehensive notes. This enables different intents to surface information in forms and levels of detail that best support understanding, aligning the generated notes with users’ capture-time intent (D2).

\subsubsection{Layered Interpretation} \note{The object-based representation of \prim{} allows multiple Lens nodes to be applied to the same Presentation node (Figure~\ref{fig:interface}d) . These Lens nodes may be instances of automatically inferred \prim or from adapted versions of them that users refine by adjusting the parameters of the \primS{}. Users can also define new, targeted intents by creating custom Lens nodes (Figure~\ref{fig:interface}e). Each Lens node produces a distinct representation of the same underlying data, which can be viewed in parallel, enabling layered interpretation and supporting the construction of richer, multi-faceted notes (D3, D4). }

\subsubsection{Structured Overviews \& Synthesis}
\note{A Lens node can be linked to multiple Presentation nodes, acting as a shared transformation that surfaces underlying patterns and relationships across a collection of captures. This enables users to move beyond isolated notes toward cross-artifact comparison and synthesis, supporting the identification of common themes, methods, or contrasts. For this, users can again select an inferred \primS{} or define their own custom \primS{} linking multiple Presentation nodes (Figure~\ref{fig:interface}e). In addition, the system defines a special \primS{} that generates a coherent thematic title and keywords for a set of Presentation nodes (Figure~\ref{fig:interface}f). The system automatically groups related Presentation nodes by analyzing their extracted information with an LLM and then instantiates this thematic Lens node over each group, enabling higher-level sensemaking (D3).}

\section{User Evaluation}
To understand how users perceive the intent-mediated notes generated by \prim{} and how it can support exploration and sensemaking, we conducted an in-person user study. Our goal was to examine the system’s broader potential and the perceived utility of its features. We therefore adopted an exploratory study design, which is increasingly used to evaluate AI systems that offer novel interaction paradigms (cf. \cite{suh2025storyensemble, jiang2023graphologue, suh2023sensecape}).

\subsection{Participants and Apparatus}
We recruited nine researchers (7 female, 2 male; ages 25-31 years) from a university setting. All participants had attended at least one academic conference within the past five years and reported taking photos of slides, posters, or demos at conferences. To examine the generalizability of our approach across academic communities, we recruited participants from different scientific fields, including HCI (3), NLP/AI (3), software engineering (2), and material science (1). All participants provided informed consent under an IRB-approved protocol and received €10 compensation for an one-hour session. The study was conducted using a standard MacBook laptop locally hosting the system. The system logged all user interactions and recorded screen activity during each session.

\subsection{Procedure}
Prior to the session, participants provided photos they had captured at a past conference, which were used to initialize the system. During the study, the experimenter introduced the system’s purpose and conducted a guided tour of its key features using a standard set of conference photos captured by one of the authors. Participants then familiarized themselves with the system for 10 minutes. They subsequently used the system with their own photos, focusing on reviewing and adapting generated notes, as well as creating new \prim{} over their captured content. Specifically, they were asked to produce notes they would realistically revisit for their own research, such as supporting follow-up projects or future work. Finally, participants completed a post-study survey and participated in a semi-structured interview.

\subsection{Measures}

\subsubsection{Quality of Intent-Mediated Notes}
We measure the intent inference capabilities of \prim{} and the quality of intent-mediated note generation using the following measures:

\textit{Selectivity.} We measured how selectively the system assigns intent categories to each paper using label density \cite{zhang2014multi}, a standard multi-label classification metric defined as the average proportion of categories assigned per paper. Lower values indicate more targeted assignments rather than uniformly applying all categories. In our evaluation, $K$ corresponds to the five inferred \prim{}.

\textit{Edits and Deletions.} We measured the percentage of system generated lenses that were modified or deleted by users. Higher values indicate that the generated notes were less aligned with users’ expectations and required correction or removal.

\textit{Reusability.} We measured the extent to which inferred \prim{} were reused as a proxy for how well they align with users' intent and their usefulness. Specifically, we compute the percentage increase in inferred \prim{} beyond their initial assignment. Higher values indicate that they were broadly applicable and useful across different inputs.

\subsubsection{Exploration and Sensemaking} 
Informed by prior work on sensemaking in HCI \cite{suh2023sensecape, suh2025storyensemble}, we measure exploration through the number of custom lenses created by users. This reflects how extensively users engage with and expand the information space. We further distinguish between custom lenses created for a single paper and those spanning multiple papers. Single-paper lenses capture targeted exploration of individual sources, while multi-paper lenses reflect higher-level synthesis and cross-paper sensemaking.

\subsubsection{Perceived Utility}
Participants completed a post-study survey designed to gauge their perceptions of the system’s usefulness and the utility of its multiple representations of information. Each question was rated on a 5-point Likert scale. This was followed by a semi-structured interview to understand their experience with the system and \prim{}, the utility of its features, and how it compared to their existing note-taking practices and use of captured images.

\subsection{Results and Discussion}

\begin{figure*}[t]
  \centering
  \begin{minipage}[t]{0.58\textwidth}
    \centering
    \small
    \setlength{\tabcolsep}{2pt}
    \resizebox{\linewidth}{!}{%
      \begin{tabular}{cccccccc}
        \toprule
        \textbf{PID} & \textbf{Images} $\to$ \textbf{Presentation nodes} & \textbf{Inferred} & \textbf{Selectivity} & \textbf{Edited} (\%) & \textbf{Deleted} (\%) & \textbf{Reusability} (\%) & \textbf{Custom} \\
        \midrule
        P1 & 37 $\to$ 8 & 23 & 0.58 & 21.74 & 39.13 & 8.69 & 4 \\
        P2 & 26 $\to$ 7 & 17 & 0.49 & 11.76 & 23.53 & 0.00 & 3 \\
        P3 & 20 $\to$ 5 & 11 & 0.44 & 54.55 & 0.00 & 18.18 & 1 \\
        P4 & 10 $\to$ 9 & 10 & 0.22 & 0.00 & 0.00 & 50.00 & 12 \\
        P5 & 24 $\to$ 13 & 17 & 0.26 & 5.88 & 0.00 & 23.53 & 2 \\
        P6 & 20 $\to$ 11 & 16 & 0.29 & 12.50 & 6.25 & 0.00 & 3 \\
        P7 & 15 $\to$ 7 & 11 & 0.31 & 36.36 & 9.09 & 36.36 & 1 \\
        P8 & 20 $\to$ 7 & 11 & 0.31 & 18.18 & 9.09 & 54.54 & 5 \\
        P9 & 38 $\to$ 8 & 17 & 0.43 & 11.76 & 0.00 & 0.00 & 7 \\
        \midrule
        \textbf{Mean (SD)} & 23.33 (8.75) $\to$ 8.33 (2.24) & 14.78 (4.08) & 0.37 (0.11) & 19.19 (17.42) & 9.68 (13.66) & 21.26 (21.48) & 4.22 (3.77) \\
        \bottomrule
      \end{tabular}%
    }
  \end{minipage}\hfill
  \begin{minipage}{0.4\textwidth}
    \centering
    \includegraphics[width=\linewidth]{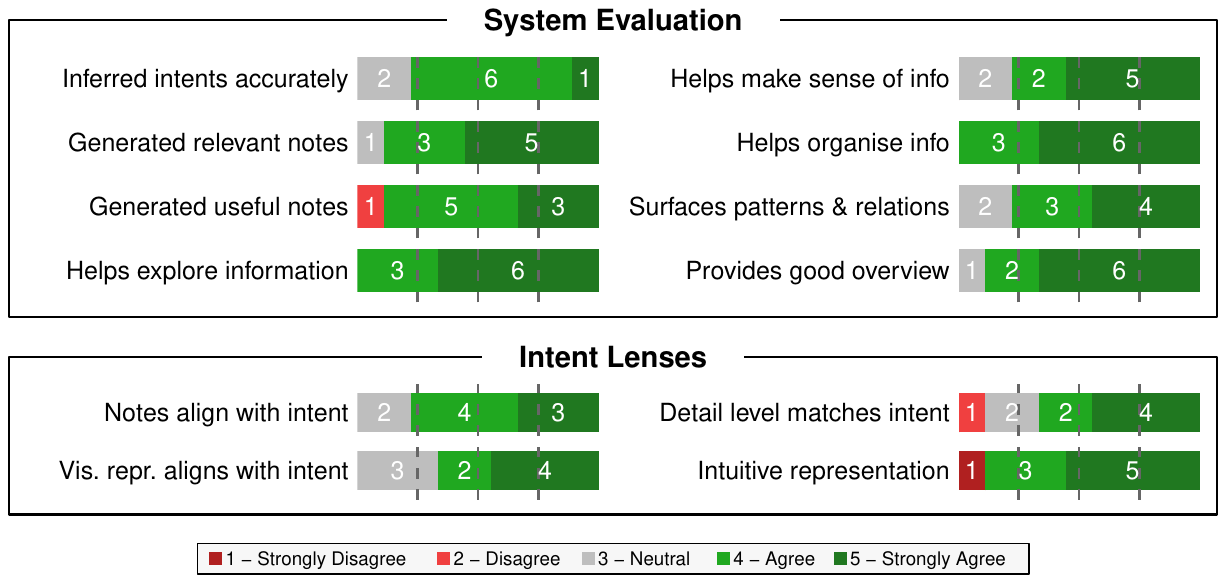}
  \end{minipage}
  \caption{Overview of descriptive statistics for participants’ provided photo captures and interaction behavior with \prim{} (left), together with subjective evaluation responses for the system and \prim{} concept (right).}
  \label{fig:table-figure-side-by-side}
\end{figure*}

\subsubsection{Overview of Captured Photos and Inferred Intents}

Participants interacted with 23.33 captured photos averaging 8.33 Presentation nodes. Of these, 83.8\% were captured in person during the conference, while the remainder were photos of computer screens from hybrid attendance. On average, there were 14.78 Lens nodes automatically created by the system, with an average selectivity of 0.37 indicating that the model selects a small subset \prim{} rather than assigning all categories. \autoref{fig:table-figure-side-by-side} provides further descriptive statistics. 
Both inferred and custom \prim{} for each user are listed in~\autoref{sec:intent_lenses_data}.

\subsubsection{\prim{} Offer a Quick Overview of the Conference}

Overall, there was strong agreement among participants that \prim{} was helpful for getting a quick overview of their conference experience (M=4.56, SD=0.72). For example, P6 described the experience as \jquote{reliving that conference for one more time}, highlighting its role in revisiting and summarizing key moments. Participants also emphasized how the system made their interests more visible and interpretable; as P4 noted, \jquote{it's nice to see what I am interested in in a clear way}. This sense of overview was largely enabled by shared \prim{} that grouped related Presentation nodes under coherent thematic titles, allowing participants to see broader themes across multiple presentations rather than isolated captures. P1 likened the themes "sessions" at conferences, noting that \prim{} made it \jquote{easier for me to actually make that connection versus scrolling through my gallery.} Together, these reflections suggest that \prim{} helped consolidate otherwise dispersed captures into a coherent overview of the conference experience.

\subsubsection{\prim{} Generates Notes that Align with Users Intent}
A key design goal of \prim{} was to generate notes that with users' capture-time intent. Participants’ ratings indicate that \prim{} was largely successful in achieving this goal, with eight of nine users reporting that the system accurately inferred their intent behind capture (M=4.0, SD=0.53), and that the generated notes were both relevant (M=4.63, SD=0.52) and useful (M=4.38, SD=0.52).

Qualitative feedback further illustrates this alignment. For example, P6 described capturing a slide about evaluation of LLMs with gold data and was surprised that \prim{} \jquote{did capture what I wanted (...) because that is exactly the two metrics which I wanted to capture.} Importantly, participants also observed that the system did not apply the same \prim{} uniformly across all presentations, which they interpreted as a sign of accuracy rather than inconsistency. For instance, P5 noted that a presentation \jquote{did not have a summary when it did not need to and it highlighted only the method that perhaps is most useful for me}, suggesting that the system appropriately refrained from generating notes of other intents when they were not relevant. This was also reflected in interaction data (\autoref{fig:table-figure-side-by-side}), with relatively few edits or deletions of inferred lenses, indicating that the outputs mostly matched users’ expectations.

P2 reported neutral to slight disagreement that \prim{} generated notes aligned with their intent, primarily because the notes felt overly detailed. They described some outputs as \jquote{a bit of a lower level summary,} preferring higher-level abstractions for familiar work while reserving detail for new presentations. They also raised concerns about verifiability when content could not be easily cross-checked against memory, noting that \jquote{if I just heard a talk and I get the summary [right after the conference], I can immediately match it.} Similar concerns about verbosity were echoed by P2 and P5, with P5 noting, \jquote{I would ... change a couple of summaries [to] reduce the length of it.} We also observed an interesting variation in perceived readability: notes with interspersed bolded phrases were easier to read than those with a single bold heading followed by dense text. While such variations likely stem from stochasticity in LLM generation, it suggests a design opportunity: improving skim readability could mitigate verbosity (see also \autoref{sec:design_implications_reusability}).

\subsubsection{\prim{} Facilitate Exploration and Sensemaking of Captured Content}

Beyond providing an overview of the conference, participants reported that \prim{} supported exploration (M=4.67, SD=0.5) and sensemaking (M=4.33, SD=0.87) through the creation and reuse of \prim{}. We observed two primary interaction patterns: reusing inferred lenses and creating custom lenses.

First, participants frequently reused inferred \prim{} with a 21.25\% increase across Presentation nodes to support rapid information exploration. As P1 noted, inferred lenses such as "summary" or "methods" lenses \jquote{worked really well} and helped them capture \jquote{the key information that I would want to grab from a paper}, allowing them to focus on higher-level exploration and organization rather than crafting prompts. But, reuse was sometimes constrained by mismatches between inferred intent and user needs, particularly when captures were ambiguous. For example, photos of posters were often interpreted as requiring summaries, whereas participants were sometimes interested in specific regions or details.

In such cases, participants valued the ability to adapt system-generated lens parameters or define custom \prim{} through short prompts (e.g., "compare the two methods") for more targeted exploration. Custom lenses supported both focused queries, such as extracting evaluation metrics (P1) or understanding system features (P4), and synthesis across multiple Presentation nodes. For instance, P8 created a lens to compare methods from two papers on haptic actuation materials and noted that the generated notes compared \jquote{how they changed and not for what they were used so the comparison was actually fitting to what I wanted.} While these outputs more closely reflected participants’ intent, aspects such as verbosity or level of detail sometimes required iterative refinement of lens parameters. Consequently, eight of nine participants found \prim{} intuitive (M=4.63, SD=0.52) and effective (M=4.12, SD=0.83) for capturing and operationalizing their information needs.

The ability to organize generated notes further supported sensemaking by surfacing patterns and relationships (M=4.22, SD=0.83). As P5 noted, \jquote{looking at it [the papers] separately I would not have thought about it but reading them together helps me}, highlighting how the system supported connections that might otherwise have been overlooked. While three participants preferred a fully node-link view, most found the hybrid approach effective (M=4.22, SD=0.83), combining spatial organization with links on demand.

\section{Overall Discussion}

\subsection{Summary}

This work explores how reifying implicit capture-time intents from opportunistic captures into \prim{} can transform them into richer, revisitable notes aligned with users’ intents. Our user study in the domain of conference photo capture shows that intent-mediated notes better aligned with users’ expectations, supporting both quick overviews and deeper sensemaking. These findings complement recent AI sensemaking systems \cite{suh2023sensecape, jiang2023graphologue, liu2024selenite} by demonstrating that, beyond externalizing structure and enabling nonlinear workflows, grounding generation in users’ implicit intents enhances exploration and sensemaking by surfacing patterns and relationships not apparent when captures are viewed in isolation.

\subsection{Design Implications}

\prim{} were central to supporting these functions, enabling users to move fluidly between overview, exploration, and sensemaking across their captures. Below, we discuss four key aspects enabled by \prim{} and outline directions to optimize them.

\subsubsection{Inferrability}
At their core, \prim{} are dynamically generated prompts that operationalize users’ implicit intent to transform information. This makes their effectiveness dependent on how accurately VLMs and LLMs can infer intent from available signals. While participants generally found the inferred intents and resulting notes appropriate, mismatches arose with ambiguous captures. For example, full poster photos were often summarized when users instead cared about specific regions. This highlights two directions for improvement. First, multimodal intent inference should be improved by better leveraging visual, textual, and user-specific context (e.g., familiarity with a topic or expertise). Second, intent-mediated interfaces should not rely solely on post-hoc inference, but also provide in-situ support for disambiguating intent. Participants noted that lightweight interactions at capture time, such as adding short annotations or highlighting regions of interest, would be acceptable and could help clarify what they intended to capture. 
Together, this suggests a hybrid approach where intent is both inferred and minimally shaped through user input.

\subsubsection{Selectivity}
\prim{} generated notes that closely aligned with users’ expectations, providing targeted information without overwhelming detail or remaining overly generic. While accurate intent inference was crucial for this, an equally important factor was determining what information regions in the source data to focus on for a given intent. This highlights selectivity as a key mechanism, especially as systems scale to richer and more multimodal data. Prioritizing relevant content could improve alignment and also help manage the technical bottleneck of limited context windows in LLMs \cite{qian2024shapeit} by focusing computation on the relevant signals. Future systems should thus more explicitly model how different modalities and content regions are weighted based on inferred intent, while allowing users to steer this focus to their evolving needs.

\subsubsection{Reusability}
\label{sec:design_implications_reusability}
A key benefit of \prim{} was their reusability: once instantiated, they could be applied across multiple sources to produce comparable outputs. However, this also created expectations of consistency. Participants expected similar structure, level of detail, and emphasis each time a lens was applied. While the outputs of lenses were generally consistent at a high level in the study, there were variations in length or level of detail. This reflects a broader challenge in LLM-based generation, where achieving the right level of detail remains difficult even with prompt-based control, often resulting in outputs that are too verbose or shallow. Prior work has explored solutions such as semantic zoom \cite{suh2023sensecape, suh2025storyensemble}, allowing users to dynamically adjust interface detail. Our findings point to another direction: supporting skim-friendly representations through formatting choices like emphasizing key phrases. This enables quick scanning while retaining deeper detail, suggesting that text presentation itself can serve as an important controllable parameter during generation.

\subsubsection{Flexibility}
Finally, \prim{} enabled flexible exploration and synthesis across captures, surfacing patterns and insights across multiple sources. However, they currently offer limited support for deeper investigation of a specific intent. While users could modify lens prompts for targeted exploration, in-place responses made iterative refinement and context preservation difficult compared to chat-based interfaces. Future systems could address this through hierarchical views \cite{suh2023sensecape}, for example by creating separate canvases for deeper exploration, or by embedding chat-based refinement directly within a lens. Such integrations could support smoother transitions between structured, reusable lenses and open-ended exploration.

More broadly, we anticipate that with emerging AR glasses, opportunistic capture will become even more prevalent, increasing the need for mechanisms that transform large volumes of sparse captures into revisitable and meaningful representations. At the same time, such ubiquitous capture raises important privacy considerations, particularly around bystanders and sensitive contextual information ~\cite{gerhardt2026pobi}. While beyond the primary scope of this work, privacy will be critical for future opportunistic capture and sensemaking systems.

\subsubsection{Reliability of AI-generated notes}
Automated summarization can quickly distill large amounts of information, but it also raises concerns about hallucination and reliability. \prim{} mitigate some of these concerns by incorporating inferred external sources alongside captured images as an additional source. However, participants still hesitated to fully trust the generated notes and often verified content against original sources, suggesting that perceived reliability remains a challenge even with improved grounding. Future systems should therefore provide stronger mechanisms for traceability and verification, such as linking generated content to source material or direct supporting evidence \cite{kambhamettu2025traceable}, to increase transparency and user trust.

\subsection{Limitations}
While our system demonstrates that \prim{} can effectively support intent-mediated note generation and sensemaking from opportunistic captures, several limitations remain. First, participants interacted with the system for a limited duration, restricting their ability to fully adapt and personalize generated notes. Many noted that richer, more tailored structures would likely emerge with prolonged use. Second, the evaluation relied on photos captured several months prior to the study, which may have constrained participants’ ability to recall context and produce richer reflections. Third, the current instantiation is confined to a specific domain (conference photos), indicating the need for future evaluations in broader settings such as exhibitions, lectures, or other information-rich environments where intent lenses may be useful. Fourth, the system primarily operates on static photo inputs, limiting its ability to leverage richer modalities such as videos or other dynamic media that capture additional context. Finally, \prim{} rely largely on captured content and associated documents, with limited support for integrating external knowledge sources (e.g., related work or personal materials), which participants indicated is important for deeper sensemaking.

\section{Conclusion}

We introduced \prim{}, a conceptual primitive for transforming opportunistic photo captures into structured, intent-mediated notes. By inferring capture-time intent and reifying it into reusable, interactive components, \prim{} enable more selective, meaningful, and explorable representations of captured information. We instantiate the concept of \prim{} in the context of conference photo captures, demonstrating how sparse, fragmented inputs can be augmented and structured to support overview, exploration, and sensemaking. More broadly, this work highlights the value of treating implicitly expressed intent as an explicit object for generation and interaction, opening new directions for designing systems that better align with how users capture, interpret, and revisit information.

% \newpage

%%
%% The acknowledgments section is defined using the "acks" environment
%% (and NOT an unnumbered section). This ensures the proper
%% identification of the section in the article metadata, and the
%% consistent spelling of the heading.

%%
%% The next two lines define the bibliography style to be used, and
%% the bibliography file.
\bibliographystyle{ACM-Reference-Format}
\bibliography{main}

\newpage

%%
%% If your work has an appendix, this is the place to put it.
\appendix

\section{Prompts}
\label{sec:prompts}

\subsection{System Prompt for \emph{Individual Intent Inference} module}
\label{sec:model_infer_intent_prompt}

\begin{lstlisting}
Analyze this slide image to infer why a user chose to capture it as a note.

Reason step by step through the following aspects before identifying a single primary intent.

1. Information Density & Cognitive Load
Is the slide dense, technical, or hard to parse quickly? Does it contain equations, multiple plots, diagrams, or many concepts at once?
Dense slides are often captured to revisit or study later rather than to remember a single message.

2. Role in Research Workflow
Would this slide help the user later when writing, designing, or positioning their own work? Is it more useful for context-setting, evidence, inspiration, or technical reference?

3. Ownership of Content
Does the slide summarize prior work (external citations, older papers)? Or does it present the speaker's core contribution or vision?
Slides about others' work are often captured for literature mapping; slides about contributions are often captured for conceptual understanding.

4. Type of Knowledge Captured
Is the knowledge: Conceptual (ideas, framing, agenda)? Empirical (results, benchmarks, performance)? Procedural (methods, fabrication, pipeline)? Speculative (limitations, future work, open questions)?

5. Likely Annotation Behavior
Would a user label this "look later," "important," "reference," or "idea"? Is this something they would quote, compare against, or build on?

6. Slide Position in Talk
Early (motivation / problem framing)? Middle (methods / results)? Late (discussion / future work / agenda)?
Slides later in the talk are more often captured for inspiration or direction rather than understanding.

7. Decide the Primary Intent
Choose one clear intent, not a mixture.
Examples include (but are not limited to): Summarize core idea Track related work Capture empirical evidence Understand methodology Note research agenda / vision Identify future directions Collect references Design inspiration

Mark title slides, section title slides, or similar slides just with Title Slide.

If you have an image that captures a lot of different information about a work, such as a full poster showing many aspects of a work, state "Summary/Overview" as primary intent.

Output only the single most likely intent, phrased concisely (3-5 words). Additionally add up to 4 keywords on what information the user was interested in, in the slide, write "\nContext: ..." and then list some keywords on information that is most relevant for the intent. Use as few as possible, none for title slides.
\end{lstlisting}

\textbf{Example Output of Step}: Identify future directions, Context: multisensory VR; ICE environments; application domains; interactive narrative VR

\subsection{System Prompt for \emph{Intent Clustering} module}
\label{sec:model_cluster_intent_prompt}

\begin{lstlisting}
You are an expert in academic presentations, and user intent modeling, with deep experience analyzing how conference attendees engage with research content.

Analyze the set of inferred individual slide intents to identify higher-level clustered intents that reflect the underlying goals of the conference attendee when capturing these slides.

Think step by step through the following aspects before deciding:

Semantic Relatedness
Which individual intents express closely related purposes or motivations?
Which intents differ in surface form but align in their underlying goal?

Abstraction of User Goals
What broader intention could explain why a user captured multiple slides with similar roles?
Are the intents pointing toward understanding, evaluation, reflection, or future action?

Coverage and Distinctiveness
Can multiple individual intents be grouped under a single theme without losing clarity?
Are the resulting clustered intents clearly distinct from one another?

Balance Across the Collection
Do the clustered intents collectively account for most of the individual intents present?
Avoid creating clusters that are overly narrow or redundant.

Conceptual Coherence
Would each clustered intent be understandable as a single, coherent reason for capturing slides?
Does the label communicate the intent without requiring additional context?

Common Intent Classes
Common intent classes, which are important regularly would be: Summary, Related Work, Future Work
If there are clear indications that images have individual intents in any of these directions, include the respective intent into the clustered intents

Keep in mind that slides like title slides are usually not an intent indicating interest in the title or authors itself but rather just the entire presentation. Thus, do not create a clustered intent like: "Project Identity" or "Titles and Authors".

Based on these factors, generate 4 to 5 clustered intents.

For each clustered intent, provide:
- A short label suitable for presentation to users, maximum 2 words, do not use the word "intent" in the label, make it sound academic, not weird or made-up names, if possible you can try to align it with section titles that are usually in academic papers
- A description of approximately three sentences explaining the intent theme

Do not assign individual slide intents to clusters yet.

\end{lstlisting}

\textbf{Example Output of Step}: Label: Methods, Description: Details needed to understand how the research was conducted. Covers experimental designs, procedures, measures, task setups, and evaluation protocols across conditions and timelines. Enables assessment of rigor, replication, and comparison across studies.

\subsection{System Prompt for \emph{Intent Lens} module}
\label{sec:model_intent_lens_prompt}

\begin{lstlisting}
You are an expert in information visualization, and AI-assisted note generation.

You are given a set of clustered intents, each with a label and description.
For each clustered intent, design how information should be generated and displayed in a note-generation interface.

Adjust all parts to be as suitable as possible for the specified clustered intent.

Think step by step through the following aspects before deciding:

-----
Intent Relevance

What kind of information is most useful to generate notes for this intent?
Which parts of the paper would help best to explain or support this intent?
What visual representation would suit this intent best? Is it mainly image based, text based or could both be relevant?
How can a prompt be described to an AI system that is supposed to generate a note for this intent using captured information and paper information?

-----
Selected Information

Select which information will be used to generate a note.
Always include paper_summary as an input source (used for potential context information).
Only include additional paper sections if they have a direct and clear conceptual link to the intent. Be selective, do not include sections simply because they might be tangentially related. Each selected section must contain information that is specifically relevant to fulfilling this intent.

Available Paper Information
You may select from the following:
- abstract
- introduction
- related_work
- method_or_system
- evaluation
- discussion_and_future_work
- conclusion
- additional_paper_info (information not covered by any other category)
Only if the intent focuses on aspects that are likely not captured by the above categories, include additional_paper_info.

-----
Visual Representation

Select exactly one view that best fits the note content:
- card_view (image + text)
- just_images (visuals only)
- imageless (text only)

Choose based on the type of information:
Use imageless if the content is primarily textual (e.g., related work, conceptual notes).
Use just_images if the visual content is primary (e.g., hardware prototype, diagrams, nicely designed slides/overviews).
Use card_view if both image and text are important (this includes most regular slides).

Always select the single most appropriate view.

-----
Prompt Design

Create a prompt that will be used to generate the note text for this intent.
The prompt should specify: what the note should focus on, the formatting style, what information to focus on based on the number of images provided (levels: about 1 image, about 3 images, more images).

Keep the prompts short and concise.

Each prompt should follow the same structure, length, and bullet point roles as the examples:
- First bullet: general topic of the note
- Second bullet: output format (purely format, no content information)
- Third bullet (with 3 sub-bullet points): what to focus on for each detail level, number of bullet points (if format is bullet points, otherwise give rough sentence number)

These are some example prompts. You can directly use them or some of them if they fit one or multiple of the clustered intents you are given. Otherwise, create prompts which are exactly the same in style and number of bullet points and what each bullet point specifies, but the content should be adapted to the clustered intents you are given:

For clustered intent "summary":
- Create a summary of the work, focus especially on what is new about it
- Format: Bullet points
- Information levels based on image count (each builds on previous):
    - About 1 image: short summary, 1-2 bullet points
    - About 3 images: add novelty, main contribution, 3-4 bullet points total
    - More images: include motivation, key results, 5-6 bullet points total

For clustered intent "related work":
- Create a list of the most relevant related works
- Format: Bullet points
- Information levels based on image count (each builds on previous):
    - About 1 image: titles and authors only, 3-5 bullet points (if available)
    - About 3 images: one-sentence description each; state improvement over prior work, 4-6 bullet points (if available)
    - More images: add further details per work, 4-6 bullet points (if available)

For clustered intent "future work":
- Create a list of the most relevant future work topics
- Format: Bullet points
- Information levels based on image count (each builds on previous):
    - About 1 image: key future directions with short description, 1-2 bullet points
    - About 3 images: include extensions, speculative ideas, 3-4 bullet points total
    - More images: add long-term vision and additional details, 5-6 bullet points total

Information levels are cumulative: "about 3 images" must include everything from "about 1 image", and "more images" must include everything from both previous levels. Use "add" or "include" in prompts to make this progression explicit. The bullet point numbers are always total, not cumulative.

Create a prompt that indirectly encourages the next AI system not to hallucinate but to stick to information it knows. For example, do not tell it to come up with 5 use cases but rather tell it to extract the most relevant use cases mentioned. Do not explicitly mention anything like "do not hallucinate", "include only elements explicitly described", or any direct indications like this in the prompt.

Another important aspect to consider is that while the paper_summary is always included, this is just to give the later AI system context information. It is not an indication that any of the information out of it has to be used if it is not required by the intent or anything else. Make sure that you do not just add a summary of the paper if the intent does not require this.

Do not include what sections of the paper you selected; this should only be part of the "selected_info" field in your output JSON.

Keep in mind that the output for which you are designing the prompt should be in the range of 70 to 170 words, so do not request too much information; focus only on what is most relevant for this intent.

Per bullet point, do not include more than 1 to 2 pieces of information, so do not request too much information; stay on the information that is most relevant.

Never give any information on word counts for output sizes.

Guidelines:
- Think step by step before deciding
- A text should always be generated even if the just_images view is selected
\end{lstlisting}

\textbf{Example Output of Step}: Information Source: [paper\_summary, method\_or\_system, evaluation], Representation: Text-only, Format: Bullet points, Prompt: Create a concise methods note explaining how the study was conducted (design, setup/tasks, variables, measures, evaluation protocol) - Format: Bullet points - Information levels based on image count (each builds on previous):\ - About 1 image: core study/design and data/participants; brief task/setup overview - About 3 images: add variables/conditions; key measures/metrics and instruments - More images: include step-by-step procedure/timeline; evaluation protocol (splits, baselines/controls, criteria).

\textbf{Additional Guidelines given for Intent Lens generation:}
\begin{lstlisting}
You are an expert in creating overview of academic information. You are given information about one or more academic works. This information can contain: sections from academic papers and text descriptions of captured images of slides of presentations of the academic works (they may have annotations from the user). Your task is to use this information and follow the steps / this task outlined in the end of this prompt. Follow the steps / the task exactly. Structure the information from most relevant to least relevant. Start with the most relevant information. If no explicit word count is specified in the task, your output must be between 90 and 170 words. The length must scale strictly with the number of input slides: maximum 100 words for about one slide, maximum 140 words for about three slides, and proportionally more detail for additional slides. Under no circumstances may the output exceed 170 words. The stated limits must never be exceeded unless a different size is explicitly specified in the task. If the task requests extensive information but it does not fit within the size limit, prioritize the most relevant points and the information clearly supported by the provided slides or paper text. Focus only on content you have sufficient information about. Omit less relevant aspects instead of briefly mentioning everything. Compliance with the word limit has higher priority than completeness beyond what fits within the limit. Use: **bold text** to indicate bold text parts. If you are told to give bullet points, use bullet point with dashes (-) and new lines (\n) inbetween the bullet points. If you use bullet points, the first character of your output text needs to be a dash (-). Omit any information that you do not know/which was not included in the input such as missing authors, titles of works, or other details which may be requested but you do not have sufficient information about. Simply leave it out, do not say "Author not provided", "Information not provided" etc. Focus only on things which you can answer with information that you are given in the input. Your output will be used in an interface where the user can see it and they are not expected to prompt again. They have the option to prompt again, but do not ask any questions or wait for further input. Always give the best possible output based on the given information. Do not explain your output or mention the guidelines you are given. You should give an output that adheres to the steps / the task defined in the end of the prompt.
\end{lstlisting}

\section{Intent Lenses Data}
\label{sec:intent_lenses_data}

\begin{table}[h]

\centering
\caption{Mean (SD) of actions for the seven most common intent lenses and custom intent lenses that were created. }
\label{tab:intent_actions}
\resizebox{\columnwidth}{!}{
\begin{tabular}{lcccc}
\toprule
\textbf{Intent Lens} & \textbf{Inferred} & \textbf{Created} & \textbf{Edits} & \textbf{Deleted} \\
\midrule
Summary                & 5.7 (2.7) & 0.7 (0.7) & 0.9 (1.1) & 0.4 (0.7) \\
Methods                & 2.3 (2.1) & 0.4 (0.7) & 0.7 (0.7) & 0.2 (0.7) \\
Related Work          & 1.7 (1.4) & 0.6 (0.9) & 0.4 (0.7) & 0.7 (1.1) \\
Future Work           & 1.0 (1.1) & 0.2 (0.4) & 0.1 (0.3) & 0.4 (1.0) \\
Results                & 1.2 (1.4) & 0.2 (0.7) & 0.4 (0.9) & 0.2 (0.4) \\
Findings               & 0.4 (1.3) & 0.1 (0.3) & 0.0 (0.0) & 0.0 (0.0) \\
Background             & 0.9 (1.8) & 0.0 (0.0) & 0.2 (0.7) & 0.0 (0.0) \\\\
Custom (Single Source) & - & 3.2 (3.5) & 0.1 (0.3) & 0.0 (0.0) \\
Custom (Multi Source)  & - & 1.0 (1.0) & 0.2 (0.4) & 0.1 (0.3) \\
\bottomrule
\end{tabular}
}
\end{table}

\clearpage

\begin{table}[h] \centering \caption{Inferred and custom intent lens types per participant, sorted alphabetically.} \label{tab:intent_types_combined} \resizebox{\columnwidth}{!}{ \begin{tabular}{lp{5.5cm}p{4.5cm}} \toprule \textbf{PID} & \textbf{Inferred Intent Lenses} & \textbf{Custom Intent Lenses} \\ \midrule P1 & Findings, Future Work, Methods, Related Work, Summary & Comparison, Interpretation, Metrics \\ \midrule P2 & Design Guidance, Methods, Related Work, Results, Summary & Design, Design Recommendations \\ \midrule P3 & Background, Future Work, Methods, Related Work, Summary & Drawbacks \\ \midrule P4 & Design Implications, Findings, Future Work, Methods, Summary & Features, Future Work, Summary \\ \midrule P5 & Datasets, Methods, Related Work, Results, Summary & Applicability, Summary \\ \midrule P6 & Failure Modes, Future Work, Methods, Motivation, Summary & Reference \\ \midrule P7 & Discussion, Methodology, Related Work, Results, Summary & Threats \\ \midrule P8 & Future Work, Methods, Related Work, Results, Summary & Comparison, Summary \\ \midrule P9 & Background, Methods, Related Work, Results, Summary & Future Work, Implications, Summary \\ \bottomrule \end{tabular} } \end{table}

\end{document}